# Intelligent Autonomous Agents are Key to Cyber Defense of the Future Army Networks

Alexander Kott

US Army Research Laboratory

**Abstract**

Intelligent autonomous agents will be widely present on the battlefield of the future. The proliferation of intelligent agents is the emerging reality of warfare, and they will form an ever growing fraction of total military assets. By necessity, intelligent autonomous cyber defense agents are likely to become primary cyber fighters on the future battlefield. Initial explorations have identified the key functions, components and their interactions for a potential reference architecture of such an agent. However, it is beyond the current state of AI to support an agent that could operate intelligently in an environment as complex as the real battlefield. A number of difficult challenges are yet to be overcome. At the same time, a growing body of research in Government and academia demonstrates promising steps towards solving some of the challenges. The industry is beginning to embrace approaches that may contribute to technologies of autonomous intelligent agents for cyber defense of the Army networks.

**A Cyber Defense Agent among Other Intelligent Things**

The landscape of possible AI applications in military seems enormously broad. However, if we were to seek the primary types of "intelligent things" (i.e., embodiments of artificial intelligence in user-relevant capabilities) most directly relevant to the future of ground warfare, we may find a rather small number of such types. In the following, I offer my vision of a pragmatic taxonomy of such intelligent entities, as they may appear on the battlefield of the mid- to long-term future (let's say in years 2035-2050). This taxonomy is not exclusive, but it does cover a large fraction of functions where AI is likely to have impact on ground combat. In this article, I intend to focus on only one of these types – the cyber defense agent – but it helps to introduce it as a member of a broader family of intelligent agents.

**Intelligent off-road ground mover**

This is a physical vehicle, a robot – whether it is legged, wheeled or tracked – intended for moving other intelligent entities (including Soldiers), supplies, munitions and weapon systems around the battlefield. Today such vehicles are largely tele-operated at low speeds, or are able to drive autonomously on well-ordered roads, or follow the leader on unimproved roads (Machi 2018).

The high lethality and dispersion of the future battlefield will make the wide use of such movers a necessity. The mover will be capable of fast and tactically appropriate movements in complex terrain, such as rough, heavy forests, mountains, and urban rubble, and possibly even climbing over obstacles using limbs.  It will self-manage its trips to charging/refueling stations and self-recharging, self-right when overturned, and autonomously load and unload.  It will plan fairly complex tasks given a general intent by the Soldier, and collaborate with other intelligent agents.

**Intelligent munition**

These physical entities will approach and defeat an adversary's asset, either by kinetic or other means. Some will resemble today's ancestors like guided artillery shells, fire-and-forget munitions, and weaponized unmanned aerial vehicles.

Future intelligent munitions will be necessitated partly by the proliferation of adversary autonomous systems. The bulk of these munitions will target the adversary munitions and information collectors. They will likely be able to conduct autonomous scene assessment and (moving) target recognition in cluttered ground environment, as well as to recognize adversary countermeasures and to perform aggressive maneuvers to avoid them. Some will be able to autonomously plan nap-of-the-earth flight, multi-munition collaboration to defeat hard targets, and collaborative allocation and pursuit of multiple authorized targets. Others will be defensive in nature such as the autonomous active protection systems (Freedberg 2016). When such intelligent munitions are used by the US, they will comply with strong constraints on autonomous and semi-autonomous weapon systems established by the US Department of Defense (Hall 2017). When the munitions are used by other countries, it is hard to predict what constraints they may comply with, if at all.

**Intelligent information collector**

Today's UAVs and UGVs collect sensor information while being largely tele-operated and following predefined waypoints. Humans make the detailed decisions about what data is to be collected, when, where, and how.

Even today, management of collection assets is burdensome for Soldiers. With the ever increasing number of such assets, future intelligent collectors will have to become broadly autonomous in formulating their paths and collection plans, based on the mission and intent provided by Soldiers. The plans could be even autonomously defined in collaboration with other collectors and based on gaps in available information.  Many of them will be small, micro-autonomous systems (Piekarski et al., 2016), that will fly, perch, and crawl in a way that minimizes their detection by the adversary. Some will be capable of fast nap-of-the-earth movements through forests and urban terrain. They will perform continuous adversarial reasoning to understand the adversary and to minimize probability of detection by the adversary. They will plan and manage their own launch and recovery, recharging and maintenance, and in general try to minimize their burden on the Soldiers.

**Intelligent information integrator and interpreter**

Today's AI models can perform functions like imagery fusion and automated detection of certain targets and patterns of activity in images and video yet, much of the collected information cannot be properly processed and interpreted. As the volume of available, collected information continues to increase, the situation will steadily get more challenging.

In the future, information integrator agents will be able to use multiple, highly dissimilar types of information in order to perform continuous recognition and interpretation of enemy and friendly activities on broad battlefield scale, along with projection of adversary upcoming activities. They will be able to collaborate with each other in a distributed operation, and communicate with Soldiers by explaining the basis for their findings, and pointing out the potential implications of the findings. They will keep up with evolving conditions and adversary by rapid learning from small number of examples. They will be capable of adversarial reasoning (inferring the adversary plans) and mindful of deception, e.g., the challenge of adversarial learning (Papernot et al., 2016).

**Intelligent COA generator and monitor**

These virtual agents will have to be far more autonomous then today's versions that support human-driven planning mainly as computerized drawing boards and maps and templates. The future battles, with high numbers of robotic assets, will acquire greater tempo and will demand detailed planning and agile execution not only for Soldiers but also the far more numerous intelligent agents. The future agents will perform largely autonomous – but collaborating with Soldiers and other intelligent agents as appropriate – preparation of plans for robotic collectors and asset movers and ongoing dynamic management of a fast-moving, robotic-heavy battle at scale with limited guidance from humans. Such an agent will operate in a distributed fashion, will collaborate closely with the intelligent information integration agents, and will conduct continuous wargaming to assess a range of alternative plans.

**Intelligent network management agent**

Today's network management tools are largely limited to centralized network controllers that display information and allow engineers to push configuration changes, often with the help of specialized scripting languages and libraries of scripts. Even today, this approach is hardly adequate for managing dynamic tactical networks and the coming decades will see networks with ever-growing complexity, diversity and fast changes in operations. Future network management agents will operate collaboratively to ensure self-forming and self-healing networks that respond to complex, large-scale disruptions, including the ability to anticipate and proactively adapt to adversarial actions. They will continually perform autonomous identification and modeling of the network, detect anomalies and perform configuration and topology changes, and manage trust.

**Intelligent cyber defense agent**

Finally, the primary topic of this article: the intelligent cyber defense agent. Today's related capabilities include fire walls, intrusion detection and alerting, and scripted removal of known malware.

In the future, just like physical robots, the cyber agents will be employed in a range of roles. Some will protect communications and information (Stytz et al., 2005) or will fact-check, filter and fuse information for cyber situational awareness (Kott et al., 2014). Others will defend electronic devices from effects of electronic warfare. These defensive actions might include creation of informational or electromagnetic deceptions or camouflage. An intelligent cyber agent will be capable of planning and execution of complex multi-step activities for defeating or degrading sophisticated adversary malware, with anticipation and minimization of resulting side effects. It will be capable of adversarial reasoning to avoid detection and defeat by adversary agents, and collaborate on planning and actions with friendly agents. In the remainder of the article, I will talk in more detail about the functions and capabilities of such agents.

The cyber agent is exceptionally important among the examples of agents I listed above. None of other agents can directly help the cyber agent survive on the battlefield of the future. At the same time, none of other agents can themselves survive without the protection of the cyber agent.

In a major conflict with a peer competitor, the friendly tactical networks will face a strongly contested environment. The sophisticated adversary will continually attack the networks and devices with cyber and electromagnetic technologies. Its capable malware – the adversary cyber agents – will, in a number of cases, penetrate and operate on the friendly devices. In other words, all intelligent agents I have described will be targets of cyberattacks. The potential that a great number of such agents will participate on the future battlefield makes cyberattacks exceptionally beneficial to the adversary, if they are successful and not effectively opposed.

Today's reliance on human cyber defenders will be untenable in the future. The proliferation of intelligent agents is the emerging reality of warfare, and they will form an ever growing fraction of total military assets (Scharre 2014). The sheer quantity of targetable friendly agents, the complexity and diversity of the overall network of entities and events, the fast tempo of robotic-heavy battle, the difficulties of centralized defense in a communications-contested environment, the relative scarcity of human Soldiers in highly dispersed operations, and the high cognitive load imposed on them by activities other than cyber defense – all make intelligent, autonomous cyber defense agent a necessity on the battlefield of the future.

In the remainder of this article, I will describe the possible functions and architecture of an intelligent autonomous cyber defense (based mainly on (Kott et al., 2018), and the limitations of today's AI (following mainly (Kott 2018)) that would need to be overcome in order to make such agents feasible and effective, and offer a few examples of today's efforts aimed at developing such agents.

**Desired Capabilities of an Intelligent Cyber Defense Agent**

In this section, I mainly follow the documents produced by a NATO Science and Technology Organization's research group on "Intelligent Autonomous Agents for Cyber Defense and Resilience", which I happen to chair. The group's objective is to help accelerate the development and transition to practical use of such intelligent agents by producing a reference architecture and a technical roadmap (Kott et al., 2018; Theron et al., 2018).

To limit the scope of the discussion, consider a single autonomous platform, such as an intelligent ground mover or an intelligent munition (such as I described earlier) with one or more computers residing on the platform, connected to sensors and actuators. Each computer contributes considerably to the operation of the platform or systems installed on the platform. One or more computers are assumed to have been compromised by the adversary malware, where the compromise is either established as a fact or is suspected.

Due to the contested nature of the communications environment (e.g., the adversary is jamming the communications, or radio silence is required to avoid detection by the adversary), communications between the vehicle and other elements of the friendly force are limited and intermittent at best. Given the constraints on communications, conventional centralized cyber defense (i.e., an architecture where local sensors send cyber-relevant information to a central location where highly capable cyber defense systems and human analysts detect the presence of malware and initiate corrective actions remotely) is often infeasible. It is also unrealistic to expect that Soldiers, even if they have direct access to the autonomous vehicle, will have the necessary skills or time available to perform cyber defense functions with respect to the vehicle.

Therefore, the cyber defense of such a platform, including its computing devices, will be performed by an intelligent, autonomous agent. The agent (or multiple agents per platform) will stealthily monitor the networks, detect the adversary agents while remaining concealed, and then destroy or degrade the adversary malware. Provisions are made to enable a remote or local human controller to fully observe, direct, and modify the actions of the agent. However, it is recognized that human control will often be impossible. Similarly, provisions are made for the agent to collaborate with agents residing on other vehicles; however, in most cases, because the communications are impaired or observed by the adversary, the agent operates alone.

To fight the adversary malware deployed on the friendly computer, the agent often has to take destructive actions, such as deleting or quarantining certain malware. Such destructive actions are carefully controlled by the appropriate rules of engagement and are allowed only on the computer where the agent resides. The actions of the agent, in general, cannot be guaranteed to preserve the availability or integrity of the functions and data of friendly computers. There is a risk that an action of the agent will "break" the friendly computer, disable important friendly software, or corrupt or delete important data. Developers of the agent will attempt to design its actions and planning capability to minimize the risk. This risk, in a military environment, has to be balanced against the death or destruction caused by the adversary if the agent's action is not taken.

The adversary malware, specifically, its capabilities and tactics, techniques, and procedures (TTPs), evolves rapidly. Therefore, the agent will be capable of autonomous learning. In case the adversary malware knows that the agent exists and is likely to be present on the computer, the adversary malware seeks to find and destroy the agent. Therefore, the agent will possess techniques and mechanisms for maintaining a certain degree of stealth, camouflage, and concealment. More generally, the agent takes measures that reduce the probability that the adversary malware will detect the agent. The agent is mindful of the need to exercise self-preservation and self-defense.

It is assumed here that the agent resides on a computer where it was originally installed by a human controller or authorized process. It is possible to envision that an agent may move itself (or move a replica of itself) to another computer. However, such propagation is assumed to occur only under exceptional and well- specified conditions, and takes place only within a friendly network—from one friendly computer to another friendly computer.

This brings to mind the controversy about "good viruses". Such viruses have been proposed and dismissed earlier (Muttik 2016). These criticisms do not apply here. This agent is not a virus, because it does not propagate except under explicit conditions within authorized and cooperative nodes. It is also used only in military environments, where most of concerns about "good viruses" do not apply.

The architecture of the agent, partly derived from the widely accepted model of Russell and Norvig (2009), is assumed to include the functional components shown in Fig. 1.

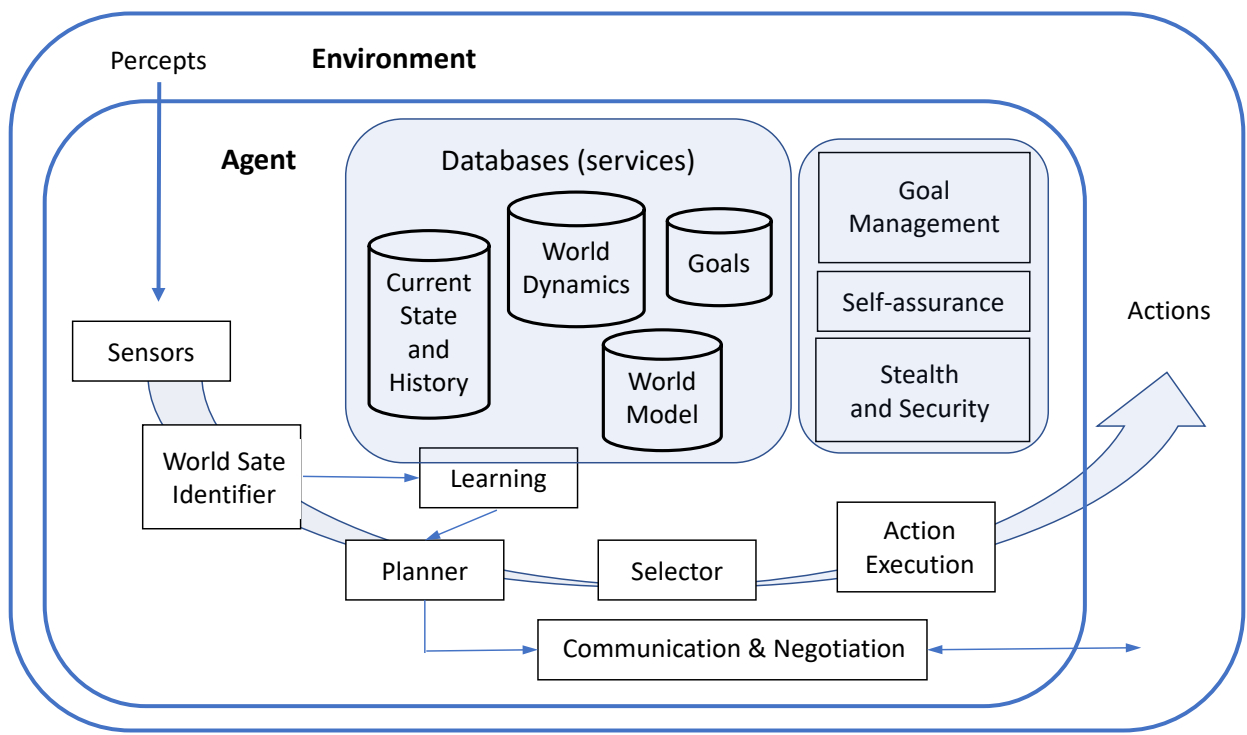

Figure 1. Functional Architecture of an Autonomous Intelligent Cyber Agent (Kott et al 2018)

**AI will be Challenged by the Complex Cyber Battlefield**

An intelligent cyber agent will have to operate on a highly complex and dynamic battlefield. Consider Fig. 2 that depicts an environment in which a highly-dispersed team of human Soldiers and intelligent agents (including but not limited to physical robots) is facing physical and cyber threats. The agents must be effective, in this unstructured, unstable, rapidly changing, chaotic, adversarial environments; they must learn in real-time, under extreme time constraints, using only a few observations that are potentially erroneous, of uncertain accuracy and meaning, or even intentionally misleading and deceptive.

Clearly, it is beyond the current state of AI to operate intelligently in such an environments – physical or cyber – and with such demands. While use of AI for battlefield tasks has been explored on multiple occasions, e.g., (Rasch et al., 2002), and AI makes things individually and collectively more intelligent, it also makes the battlefield more difficult to understand and manage. Agents and Soldiers have to face a much more complex, and unpredictable world where intelligent agents have a mind of their own and perform actions that may appear inexplicable to the humans. Direct control of such intelligent agents by humans becomes impossible or limited to cases of whether to take a specific destructive action.

An intelligent cyber agent will need to deal with a world where sheer number and diversity of cyber objects will be enormous. The number of connected computing devices, for example within a future Army brigade, is likely to be several orders of magnitude greater than in current practice. This, however, is just the beginning. Consider that computing devices belonging to such a brigade will inevitably interact – willingly or unwillingly – with devices owned and operated by other parties, such as those of the adversary or owned by the surrounding civilian population. If the brigade operates in a large city, where each apartment building can contain thousands of devices, the overall universe of connected items grows to enormous numbers. A million devices per square kilometer is not an unreasonable expectation.

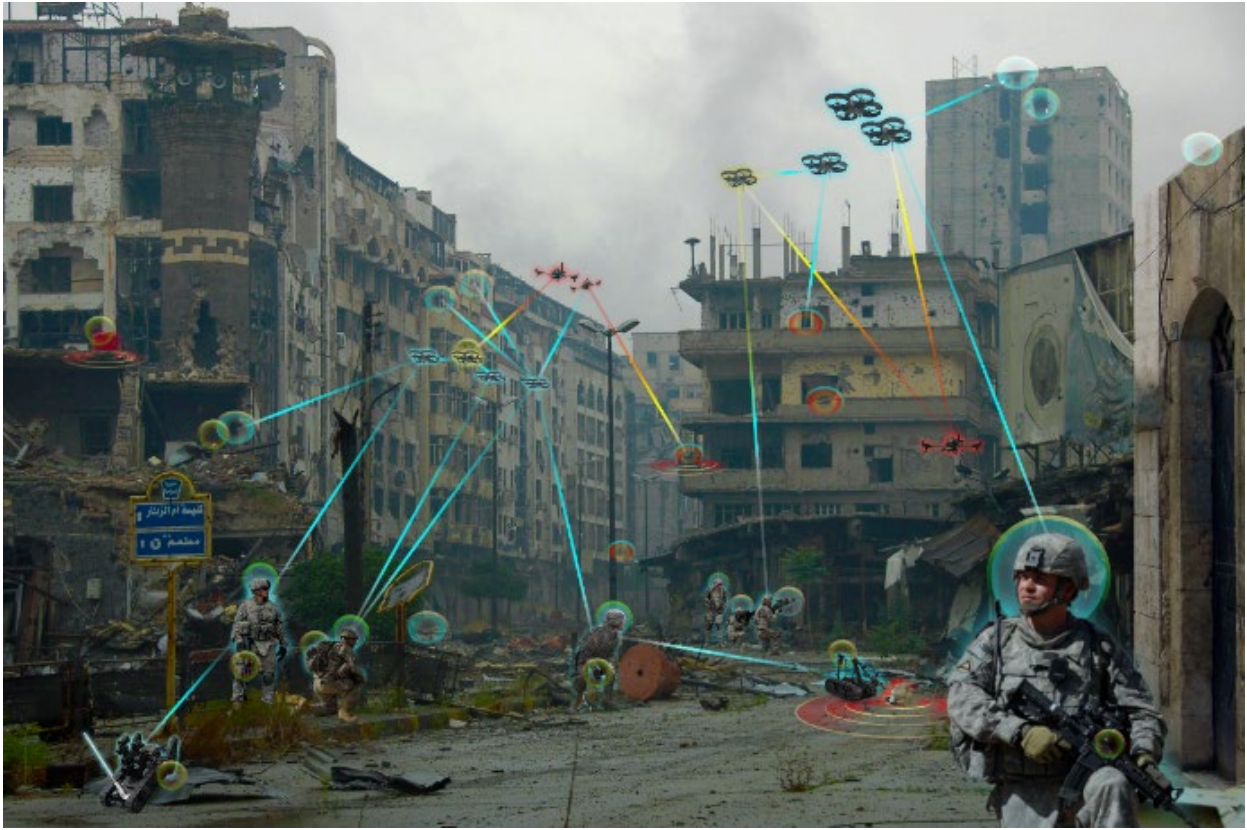

*Figure 2. An intelligent cyber agent will operate in extremely complex, challenging environment: unstructured, unstable, rapidly changing, chaotic, adversarial and deceptive.*

The above scenario also points to a great diversity of devices within the environment of the intelligent cyber agent. Devices will come from different manufacturers, with different designs, capabilities, and purposes, configured or machine-learned differently, etc. No individual agent will be able to use pre-conceived (pre-programmed, pre-learned, etc.) assumptions about behaviors or performance of other agents or devices it meets on the battlefield. Instead, behaviors and characteristics will have to be learned and updated autonomously and dynamically during the operations. This includes humans, and therefore the behaviors and intents of humans, such as friendly warfighters, adversaries, and civilians and so on, will have to be continually learned and inferred.

And yet, Machine Learning, an area that has seen a dramatic progress in the last decade, must experience major advances in order to become relevant to the real battlefield. Learning with very small number of samples is clearly a necessity in an environment where the adversary and friends change tactics continuously, and the environment itself is highly fluid, rich with details, dynamic and changing rapidly. Furthermore, very few if any of the available samples will be labelled, or at least not in a very helpful manner.

Some samples may be misleading in general, even if unintentionally (e.g., an action succeeds even though an unsuitable action is applied) and the machine learning algorithms will have to make the distinction between relevant and irrelevant, instructive and misleading. In addition, some of the samples might be a product of intentional deception by the adversary. In general, issues of Adversarial Learning (Papernot et al., 2016) and Adversarial Reasoning (Kott and McEneaney 2006) are of great importance.

Yet another challenge that is uniquely exacerbated by battlefield conditions are constraints on the available electric power and computing power. Today, most successful AI relies on vast computing and electrical power resources including cloud-computing reach-back when necessary. The battlefield AI, on the other hand, must operate within the constraints of edge devices. This means that computer processors where the intelligent cyber agent resides must be relatively lights and small, and as frugal as possible in the use of electrical power. One might suggest that a way to overcome such limitations on computing resources available directly on the battlefield would be to offload the computations via wireless communications to a powerful computing resource located outside of the battlefield. Unfortunately, this is not a viable solution, because the adversary's inevitable interference with friendly networks will limit the opportunities for use of reach-back computational resources.

**Current Efforts towards Development of Intelligent Cyber Agents**

In spite of the profound challenges, foundational capabilities are gradually emerging that would contribute to an autonomous intelligent cyber defense agent I describe here. For example, I already mentioned the NATO research group (initiated in 2016 under the title IST-152-RTG "Intelligent Autonomous Agents for Cyber Defense and Resilience.") The group is in the process of conducting focused technical analysis to produce a first-ever reference architecture and technical roadmap for autonomous cyber defense agents (Kott et al., 2018).

The group's future plans include the study of use cases that could serve as a reference for the research, as would lead to clarifying the scope, concepts, functionality, and functions' inputs and outputs of such an intelligent agent. The initially assumed architecture will be refined by drawing further lessons from the case studies. In addition, the group is working to identify and demonstrate selected elements of such capabilities, which are beginning to appear in academic and industrial research.

Based on the analysis of the proposed architecture and available technological foundation, the group is developing a roadmap towards initial yet viable capabilities. The first phase of the roadmap will include the development of knowledge-based planning of actions, the execution functionality, elements of resilient operations under attack, and adaptation of the prototype agent for execution of a small computing device. This phase would culminate in a series of Turing-like experiments that would evaluate the capability of the agent to produce plans of remediating a compromise, as compared to experienced human cyber defender.

The second phase would focus on adaptive learning, the development of a structured world-model, and mechanisms for dealing with explicitly defined, multiple and potentially conflicting goals. At this stage, the prototype agent should demonstrate the capability, in a few self-learning attempts, to return the defended system to acceptable performance after a significant change in the adversary malware behavior or techniques and procedures.

The third phase would delve into issues of multi-agent collaboration, human interactions, and ensuring both the stealth and trustworthiness of the agent. Cyber-physical challenges may need to be addressed as well. This phase would be completed when the prototype agents are able to successfully resolve a cyber compromise that could not be handled by any individual agent.

Relevant research in academia and in the Army research organizations is growing. Let me mention a few examples. Deployment of an intelligent cyber defense agent on an edge device with limited computational power requires very light yet effective packet analysis capability. Researchers at the US Army Research Laboratory developed such extremely lightweight intrusion detection prototype (Chang et al., 2013) and a similarly lightweight malware traffic classification algorithm that uses continuous machine learning (Ken and Harang 2017). Approaches are also emerging that would enable an intelligent agent to autonomously patch a software on a lightweight device once a vulnerability in that software is detected (Azim et al., 2014).

In cases when a cyber agent defends an agent with physical functions, such as an intelligent ground mover, or a collector, detection and remediation of a cyber-physical attack are particularly important. In that respect, an interesting example is the research at Purdue University (Fei et al., 2018). An autonomous agent was installed on a quadcopter. A series of attacks were then launched by embedding malicious code in the control software and by altering the vehicle's hardware with the specific targeting of sensors, controller, motors, vehicle dynamics, and operating system. Experimental results verified that the agent was capable of both detecting a variety of cyber-physical attacks, while also appropriately taking over the control system in order to recover from such attacks.

Deception and related techniques are among the most effective actions that an intelligent cyber agent can take to defend a system against a malware, while remaining undetected by the malware and its command and control operators. An example of research in that direction is described in (Asaleh et al., 2017) where an agent performs dynamic analysis of the detected malware and then plans and executes several types of deceptive actions depending on the behavior and intents of the malware. The malware remains unaware that it is being deceived. Similarly, a commercial product from Attivo Networks (Woodard 2017) helps achieve network security by luring, engaging and trapping threats and malware from infected clients and servers in the user network, data center, cloud, and SCADA/ICS network.

Speaking of commercial products, the industry is rapidly growing and evolving a space of products called Endpoint Protection Platforms (EPP) and Endpoint Detection and Response (EDR). These deserve a separate discussion (Gartner 2018). They are clearly driven by some of the same motivations, and would depend on some of the same technology advances that I discuss in this article. It is likely, however, that such commercial solutions will continue to rely on assured access to a centralized server or cloud support, and for this reason will prefer to limit the autonomy of the host-based agent.

**Conclusions**

Intelligent autonomous agents are a key type of intelligent entities that will be widely present on the battlefield of the future. The proliferation of intelligent agents is the emerging reality of warfare, and they will form an ever growing fraction of total military assets. By necessity, intelligent autonomous cyber defense agents are likely to become primary cyber fighters on the future battlefield. Indeed, today's reliance on human cyber defenders will be untenable in the future. The reasons include the sheer quantity of targetable friendly agents, the complexity and diversity of the overall network of entities and events, the fast tempo of robotic-heavy battle, the difficulties of centralized defense in a communications-contested environment, the relative scarcity of human Soldiers in the highly dispersed operations, and the high cognitive load imposed on them by activities other than cyber defense.

Initial explorations have identified the key functions, components and their interactions for a potential reference architecture of such an agent. However, it is beyond the current state of AI to support an agent that could operate intelligently in an environment as complex as the real battlefield. A number of challenges are yet to be overcome. The agents must be effective in an unstructured, unstable, rapidly changing, chaotic, adversarial environments; able to learn in real-time and under extreme time constraints, using only a few observations that are potentially erroneous, of uncertain accuracy and meaning, or even intentionally misleading and deceptive. At the same time, a growing body of research in the U.S. Government and academia demonstrates promising steps towards solving some of these challenges, and the industry is beginning to embrace approaches that may contribute to technologies of autonomous intelligent agents for cyber defense of the Army networks.

## Disclaimers

The views expressed in this paper are those of the author and not of his employer; they are not to be construed as an official Department of the Army position unless so designated by other authorized documents. Citation of manufacturer's or trade names does not constitute an official endorsement or approval of the use thereof.